\documentclass[]{akari2017}

\usepackage{here}
\usepackage{graphicx}
\usepackage{sidecap}

\shorttitle{Multi-wavelength observations of planet forming disks}
\shortauthors{I.\ Kamp, et al.}

\begin{document}

\title{Multi-wavelength observations of planet forming disks: Constraints on planet formation processes} 



\correspondingauthor{Inga Kamp}
\email{kamp@astro.rug.nl}

\author{Inga Kamp} 
\affiliation{Kapteyn Astronomical Institute, University of Groningen, The Netherlands} 
\author{Stefano Antonellini} 
\affiliation{Kapteyn Astronomical Institute, University of Groningen, The Netherlands and Astrophysics Research Centre, School of Mathematics and Physics, Queen’s University Belfast, University Road, Belfast BT7 1NN, UK} 
\author{Andres Carmona} 
\affiliation{IRAP, Universit\'e de Toulouse, CNRS, UPS, Toulouse, France} 
\author{John Ilee} 
\affiliation{Institute of Astronomy, University of Cambridge, Madingley Road, Cambridge CB3 0HA, UK} 
\author{Christian Rab} 
\affiliation{Kapteyn Astronomical Institute, University of Groningen, The Netherlands} 



\begin{abstract}
Our understanding of protoplanetary disks has greatly improved over the last decade due to a wealth of data from new facilities. Unbiased dust surveys with Spitzer leave us with good constraints on the dust dispersal timescale of small grains in the terrestrial planet forming region. In the ALMA era, this can be confronted for the first time also with evolutionary timescales of mm grains in the outer disk. Gas surveys in the context of the existing multi-wavelength dust surveys will be a key in large statistical studies of disk gas evolution. Unbiased gas surveys are limited to ALMA CO submm surveys, where the quantitative interpretation is still debated. Herschel gas surveys have been largely biased, but [O\,{\sc i}]\,63\,$\mu$m surveys and also accretion tracers agree qualitatively with the evolutionary timescale of small grains in the inner disk. Recent advances achieved by means of consistent multi-wavelength studies of gas AND dust in planet forming disks reveal the subtleties of the quantitative interpretation of gas surveys. Observational methods to determine disk masses e.g. from CO submm lines require the knowledge of the dust properties in the disk. Understanding not only the gas evolution, but also its chemical composition will provide crucial input for planet formation models. Kinetic chemical results give profoundly different answers than thermodynamic equilibrium in terms of the C/O ratios as well as the water ice/rock ratios. Again, dust has a key impact on the chemical evolution and composition of the gas. Grain growth for example affects freeze-out processes and strongly increases the cosmic ray induced UV field.  
\end{abstract}


\keywords{Solar system and planetary formation; Star formation}

\setcounter{page}{1}



\section{Introduction} 
\label{sec:intro} 

Planet formation must be an extremely efficient process accompanying star formation. We know that every young star is born with a thin rotating disk of dust and gas around it. And we know from Kepler surveys that almost every star in the Milky Way is likely hosting planets. Hence, the term planet forming disks may be more appropriate than the commonly used term protoplanetary disks. 

Comparing planet forming disk studies to evidence from our own Solar System provides interesting clues to understand when planets actually start to form. \citet{Connelly2012} revise the chronology of Calcium-Aluminium rich inclusions (CAIs) and chondrules formed very early in the Solar System showing that their formation corresponds rather to the class 0/I stage in planet forming disks. However, some ambiguity remains in matching the Solar System lookback time to that of star formation (when is time zero in star formation?). So, do the first steps towards forming larger solid bodies occur in class 0/I, or only in the class II stage?

A second key question is how planets form. If disks form very early already in the class 0/I stage like ALMA observations indicate now \citep{Tobin2012,Murillo2013,Segura-Cox2016,Tobin2016,Perez2016}, would they be initially very massive and gravitationally unstable? Is there a clear answer in favor of core (pebble) accretion or gravitational instabilities, or do both modes of planet formation exist --- possibly at different evolutionary stages or distances from the central star?

Key processes that have been proposed in the context of the core accretion scenario are (1) grain growth, (2) dust radial migration, (3) dust vertical settling and (4) dust concentrations in disks \citep[see recent review by][]{Birnstiel2016}. Some of these processes have been clearly demonstrated observationally. Dust growth has been seen first indirectly through spectral energy distributions \citep{DAlessio2001} and then through  spatially resolved disk images in multiple bands \citep[e.g.][]{Joergensen2007,Ricci2010}. Radial migration has been reported as one explanation of multi-wavelength continuum imaging of TW Hya \citep{Andrews2012}. Vertical settling has been demonstrated to be very efficient for mm-sized grains using the high spatial resolution ALMA HL Tau images \citep{Pinte2016}. Also efficient local concentration of grains has been revealed by comparison of mm observations of large dust grains (ALMA, VLA) with scattered light observations tracing small dust (Subaru/HiCIAO, VLT/SPHERE). Observations show for example dust traps for IRS48 \citep{vanderMarel2013}, concentric rings in thermal emission for HL Tau \citep{ALMApartnership2015} and spirals and arcs in scattered light \citep{Hashimoto2012,Benisty2015,Follette2015}.

All of these processes fit into the core accretion scenario. However, gravitational instabilities in early evolutionary stages of disks cannot be ruled out unless we understand better the physical conditions of the gas in the disk and can evaluate e.g. the Toomre criterium directly from local gas conditions (surface densities, gas temperatures). In the following, three selected topics will be discussed in this context: (1) Disk dust and gas evolution and dispersal, (2) Multi-wavelength disk studies in the context of gas masses, and (3) chemical evolution in disks in relation to exoplanet (atmosphere) composition.

\section{Disk evolution and dispersal}
\label{Sect:DiskEvolDisp}

Several processes can affect the disk evolution and dispersal and they act at different radial scales as well as on different timescales. Examples are accretion, multiplicity, photoevaporation (stellar FUV and X-rays, external stars), grain growth and migration, and planet formation. Recent PP\,{\sc vi} book chapters discuss the processes as well as their relation to observations \citep{Alexander2014,Espaillat2014}. The following sections present a short summary of the key findings from unbiased dust surveys at different wavelengths as well as more recent gas surveys.

\subsection{Dust surveys}

The Spitzer Space Telescope has carried out numerous unbiased dust surveys and studied the fraction of stars with near-IR excess as a function of cluster age \citep[e.g.][]{Evans2003,Carpenter2006,Hernandez2007,Meyer2008,Hernandez2008}. All studies converged on an inner warm dust dispersal timescale of a few Myrs. \citet{Takita2010} reported that the AKARI/IRC point source catalogue recovered indeed the bright end of the Spitzer T Tauri star sample from \citet{Rebull2010}, albeit in only two bands compared to Spitzer's four IRAC and three MIPS bands. Due to its much wider area coverage (all sky survey) they also discovered 28 new T Tauri candidates that need to be followed up. AKARI was at wavelengths lower than $50~\mu$m a factor of few to 10 less sensitive than Spitzer. In addition, many star forming regions in the solar neighborhood were already surveyed with Spitzer. 

More recently, \citet{Ribas2014,Ribas2015} have performed a large statistical analysis of combined photometric data from several catalogues including Spitzer and WISE. Triggered by earlier results for a mass dependent dust evolution in protoplanetary disks by \citet{Carpenter2006}, \citet{Kennedy2009}, and \citet{Yasui2014}, they used data from 22 nearby young star forming regions and showed that disks evolve faster around higher mass stars ($M\!\geq\!2$~M$_\odot$), but the fraction of “evolved” disks (defined as disks without excess at 5.8 and 8~$\mu$m) is roughly constant over 10~Myr. Explanations for faster dispersal of disks around higher mass stars could be differences in photoevaporation and/or accretion rates.

While more theoretical investigations on disk dispersal at the inner edge are underway \citep[see][for a recent review]{Gorti2016}, large surveys at longer wavelengths provide disk lifetimes for the mm-sized dust in the outer disks. \citet{Williams2013} and \citet{Andrews2013} pioneered this using SMA SCUBA-2 data and dealing with sample biases. More recently ALMA surveys expanded the evolutionary picture adding in more clusters at various ages \citep{Carpenter2014,Pascucci2016,Ansdell2016,Ansdell2017}. The level of submm emission in young clusters seems rather constant over 3 Myr. At later ages (5-10 Myr, e.g. Upp Sco), disks have substantially lower mm dust masses. This could be tentative evidence for inside-out dispersal, but more studies and larger surveys are required, especially also to characterize the influence of the environment and to distinguish between low and high mass stars like in the near- to mid-IR surveys.

\subsection{Gas surveys}

The first gas surveys were carried out using CO submm lines \citep[e.g.][]{Dent2005,Schaefer2009}; however, sample sizes were limited to less than 60 targets and mostly biased towards systems detected in the continuum at IR/submm wavelengths or very young systems (high H$\alpha$ emission). The detection rate in CO J=3-2 was $\sim\!20-30$\%. The results suggest a correlation between the CO line intensity and the fractional total IR excess (as derived from the SED). \citet{Fedele2010} used VLT/VIMOS to identify the fraction of accreting stars (through H$\alpha$ equivalent width and 10\% width) in a number of clusters and compare this to the Spitzer near-IR excess evolutionary timescale. The gas accretion is taken here as evidence for the presence of inner gas-rich disks. The gas accretion timescale follows a very similar time dependence as the near-IR dust excess with possibly a slightly shorter characteristic timescale: 2.3~Myrs for accretion versus 2.9~Myrs for the dust; note however that errors on the fractions of accreting/IR excess stars in each cluster are not folded in here.

A large observing program on CO ro-vibrational lines has been carried out with VLT/CRIRES \citep{Brown2013}. The sample comprises 69 protoplanetary disks at different evolutionary stages and in different star forming regions spanning spectral types M to B. They find a large diversity of line profiles, including broad single peaked non extended emission (with asymmetric spectro-astrometric signal), which could indicate a disk wind \citep{Pontoppidan2011}. In low mass systems, transition disks show Keplerian CO profiles that tend to be narrower than non-transitional disks, potentially indicating that the gas resides further out. Recently, \citet{Banzatti2015} show that the CO line profiles (combined broad and narrow component versus single components) and excitation could be explained with a gap opening evolutionary scenario in which the single component profiles predominantly occur for disks with dust gaps. The lack of high velocity wings in the CO lines of transition disks indicates that there is a lower column of molecular gas within few au in their disks.

The largest gas survey with Herschel was carried out by the Open Time Key Program GASPS \citep{Dent2013}. It searched primarily for [O\,{\sc i}]\,63\,$\mu$m line emission in $\sim\!250$ young stars from seven star forming regions/associations in the age range $0.3-30$~Myr plus a sample of Herbig Ae/Be stars. Of the targets with dust masses in excess of $10^{-5}$~M$_\odot$, 84\% are detected in the [O\,{\sc i}] line. The line detection rate per star forming region/association in the GASPS sample clearly decreases with age similar to what is seen for the dust. Beyond 10-20~Myr, none of the stars is detected in [O\,{\sc i}].

The recent ALMA surveys of several star forming regions \citep[Lupus, $\sigma$~Orionis, Chamaeleon~{\sc i}, by][]{Ansdell2016,Ansdell2017,Long2017} also contain the $^{12}$CO, $^{13}$CO and C$^{18}$O isotopologues lines. The isotopologue lines however are only detected in a subsample of the objects (typically less than 30\%). First analysis suggests evidence for strong gas depletion (factor 10-100) already at the stage of 1-3~Myrs \citep{Ansdell2016}, but we will come back to this in the next section.

\section{Multi-wavelengths combined gas and dust studies}

\begin{figure}[!ht]
\begin{center}	
\vspace*{-3mm}
\resizebox{0.65\hsize}{!}{
	\includegraphics*{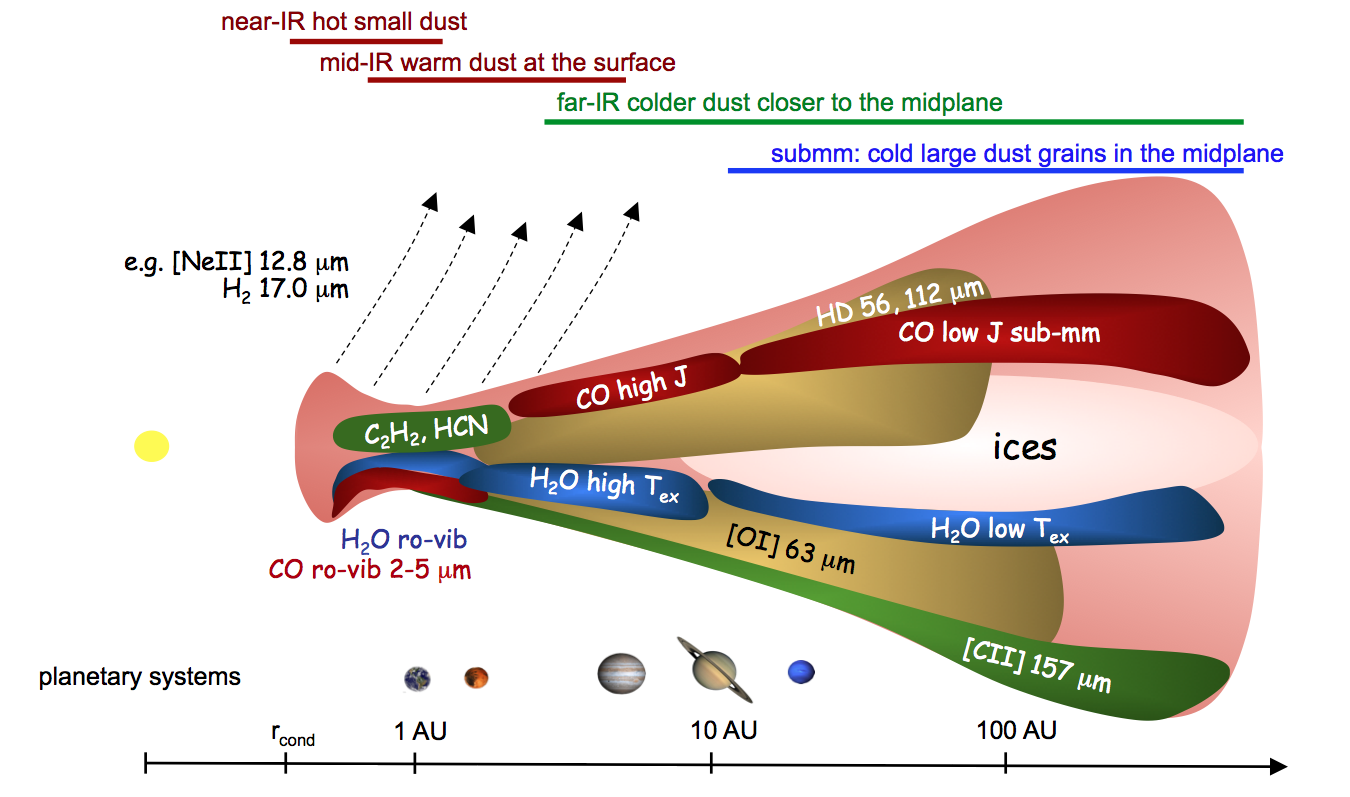}	
	}
	\caption{Sketch of a typical protoplanetary disk and the various line emitting regions in comparison to the extent of our Solar System.
	}\label{fig:disk-sketch}
\end{center}
\vspace*{-5mm}
\end{figure}

Planet forming disks span a wide range of physical conditions ranging from dense hot planetary atmosphere conditions (around $\sim\!1$~au) to cold low density interstellar medium conditions (around $\sim\!100$~au). Figure~\ref{fig:disk-sketch} shows that we need a large variety of wavelengths to obtain a full inventory of the gas physical state and chemical composition across the entire disk. Very generally, the inner disk is best studied at near-IR wavelengths, while the outer disk is best studied at submm wavelengths. The proposed SPICA mission (Roelfsema et al.\ submitted) will cover the full wavelength range from mid-IR to far-IR simultaneously and thereby provide access to a large number of emission lines ranging in excitation from roughly thousand Kelvin down to 100~Kelvin. Such capabilities are essential in the future for large unbiased gas surveys of young planet forming disks.

\subsection{Disk masses from CO submm lines}

\citet{Williams2014} proposed to use two CO isotopologues lines ($^{13}$CO and C$^{18}$O J=2-1) to deduce disk gas masses from observations. Their diagnostic diagram was later refined by \citet{Miotello2016} taking into account also selective photodissociation of CO. The method was applied in large ALMA snaphot surveys of CO isotopologue lines in star forming regions to obtain disk gas masses complementing the dust masses derived from the submm continuum \citep[e.g.][]{Ansdell2016}. Assuming a CO/H$_2$ ratio of $10^{-4}$, the detected gas masses are mostly below a Jupiter mass and the gas-to-dust mass ratios are well below the canonical 100. 

\begin{figure}[!ht]
\begin{center}	
\vspace*{-4.mm}
\resizebox{1\hsize}{!}{
	\includegraphics[width=0.5\textwidth]{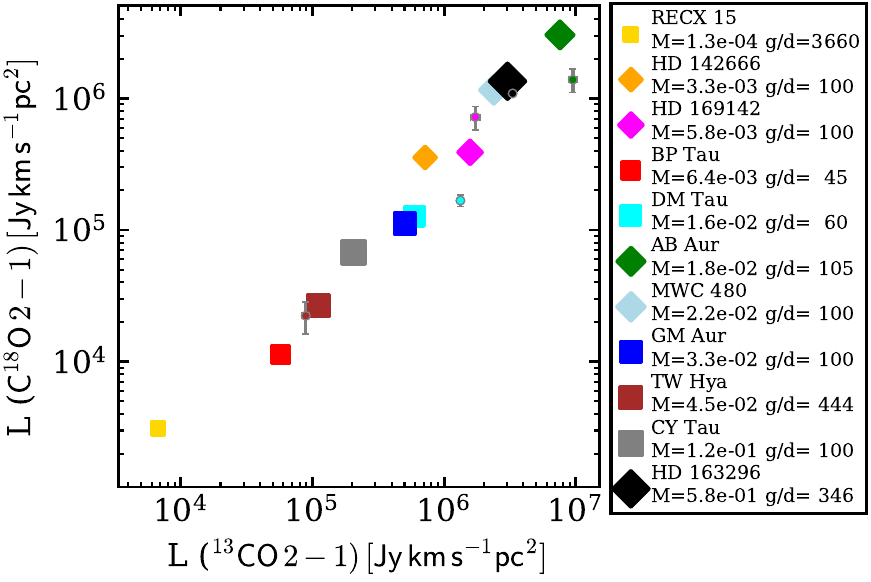}	
	\includegraphics[width=0.49\textwidth]{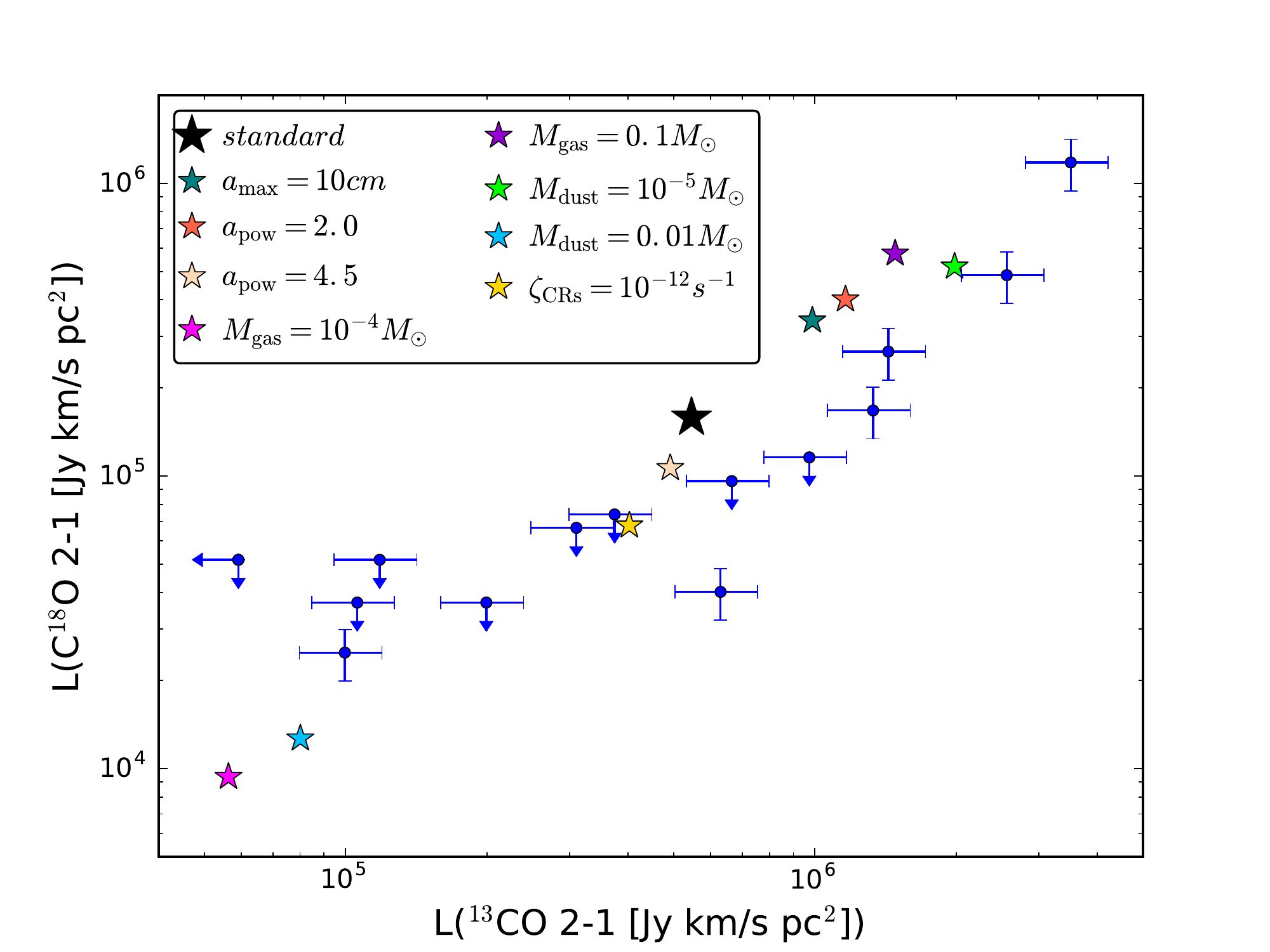}	
	}
	\caption{Left: C$^{18}$O J=2-1 versus $^{13}$CO J=2-1 line luminosities from DIANA multi-wavelengths disk models for 11 objects. The legend on the right shows the color code for each object, the total disk gas mass and the gas-to-dust mass ratio (g/d) for the fitted disk models around T Tauri stars (squares) and Herbig stars (diamonds). The size of the symbol scales with the disk mass. For those objects that do have both line fluxes detected, we also show the observed fluxes as a filled circle with the same color (5 out of 11 objects). Right: C$^{18}$O J=2-1 versus $^{13}$CO J=2-1 line luminosities for a sample from \citet[][blue circles, arrows denote upper limits]{Williams2014}. Colored symbols are the results from our T Tauri disk model series where a single disk parameter is modified with respect to the standard model \citep{Antonellini2015}. The specific parameters and values for each model are indicated in the legend. Models with different $M_\mathrm{gas}$ have the same $M_\mathrm{dust}$ (10$^{-4}$M$_{\odot}$), while models with different $M_\mathrm{dust}$ have the same $M_\mathrm{gas}$ (0.01M$_{\odot}$).
	}\label{fig:DIANA-COgasmasses}
\end{center}
\vspace*{-6.mm}
\end{figure}

Over the last years, the DIANA\footnote{Details on the DIANA FP7 SPACE project, a link to the models as well as the database can be found at http://dianaproject.wp.st-andrews.ac.uk.} team has developed a standardized approach to fit multi-wavelength observations of protoplanetary disks for dust and gas self-consistently \citep{Woitke2016}. The results of systematic variations of several disk parameters, e.g.\ the dust size distribution, strength of settling, the shape of the tapering edge as well as the disk flaring in the outer disk all affect this specific isotopologue line ratio by $\sim\!50$\%. Especially the shape of the tapered edge affects the individual line fluxes differently and for $^{12}$CO up to a factor three \citep{Woitke2016}. Based on the currently available 11 detailed disk models fitted to the multi-wavelength observations (gas and dust) of those targets, Fig.~\ref{fig:DIANA-COgasmasses} (left panel) shows that also quite massive disks can populate the lower left corner of the diagram. The degeneracy of the proposed method with especially dust properties and disk shape remains large. This is also illustrated in Fig.~\ref{fig:DIANA-COgasmasses} (right panel), where we show the line luminosities from our T Tauri standard model\footnote{The basic disk parameters are $M_{\rm gas}\!=\!10^{-2}$~M$_\odot$, $M_{\rm dust}\!=\!10^{-4}$~M$_\odot$, inner radius $R_{\rm in}\!=\!0.1$~au, tapering radius $R_{\rm taper}\!=\!200$~au, surface density power law index $\epsilon\!=\!1.0$, flaring angle $\beta\!=\!1.13$, $a_{\rm min}\!=\!0.05$~$\mu$m, $a_{\rm max}\!=\!1000$~$\mu$m and a power law index of 3.5 for the astronomical silicate dust size distribution.} \citep[black star,][]{Antonellini2015} and how they change if a single parameter is varied (colored stars, see legend). It is interesting to note that changing dust properties moves the CO lines along the same relation as disk gas mass or disk dust mass; a gas-to-dust mass ratio of 1 with a disk mass of $10^{-2}$ and $10^{-4}$~M$_\odot$ lead to the same low CO line luminosities (magenta and blue star). In the future, multi-band ALMA observations that spatially resolve the disks become available for many surveys and can help to break remaining degeneracies and provide considerable improved gas masses.

\subsection{Disk masses from HD lines}

After the first detection of HD with the Herschel/PACS instrument in the disk around TW Hya \citep{Bergin2013}, two more sources have been detected \citep[][DM Tau and GM Aur]{McClure2016} and the HD line fluxes have been used to constrain the disk gas masses; the reported masses are generally higher or at the upper end of the CO derived gas masses. Recently, \citet{Trapman2017} calculated the HD lines at 112 and 56~$\mu$m from a grid of disk models to quantify the uncertainties of gas mass derivations and also the capabilities of future observatories such as SOFIA/HIRMES and SPICA. Apart from the disk gas mass, the HD line fluxes are sensitive to the vertical disk structure, and the dust parameters such as the fraction of large dust grains and settling of dust. While SEDs alone are often too degenerate, some of these dust parameters can be constrained in the future by ALMA observations or by a combination of gas tracers such as dominant far-IR cooling lines. Overall, the HD lines largely trace warm gas inside $\sim\!100$~au, very similar to the [O\,{\sc i}]\,63~$\mu$m line \citep{Kamp2010}. For the three disks detected in HD with Herschel, \citet{Trapman2017} find that the line fluxes and 870~$\mu$m continuum fluxes are consistent with a gas-to-dust mass ratio of 100. 

While SOFIA/HIRMES will detect the brightest disks in HD with sensitivities that are slightly better than Herschel/PACS, deep surveys with SPICA/SAFARI can detect the line in disks down to $10^{-4}$~M$_\odot$. This will open for the first time the opportunity for an unbiased disk gas mass survey at the same sensitivity as dust surveys. 

\subsection{Gas column densities: The local amount of material to form planets}

Efforts to determine global gas-to-dust mass ratios in disks can be important for large population evolution studies. However, in the context of the recently detected plethora of disk substructures, such a global quantity will be limited in use. An alternative way of quantifying the local planet formation capacity could be local dust size information as well as local gas column densities in a key region for planet formation, i.e.\ the inner few to 10~au. The surface density structure in the inner disk and/or inside the gap for transitional disks can also be key in distinguishing the various mechanisms that can lead to inner disk clearing (see Sect.~\ref{Sect:DiskEvolDisp}).

Two observational approaches have been used: CO ro-vibrational line profiles from near-IR high resolution spectroscopy (e.g.\ VLT/CRIRES and Keck/NIRSpec) and spatially and spectrally resolved CO submm line data from ALMA. In both cases, the more optically thin isotopologues play a crucial role to obtain a direct column density measurement. \citet{Carmona2014} show from CO ro-vibrational line profiles in combination with a large set of multi-wavelength data, that the disk around HD135344B hosts still gas inside the submm cavity, but that the surface density power law is inverted with respect to the one in primordial disks. This is consistent for example with hydrodynamical simulations of a massive planet opening a gap in a disk \citep[e.g.][]{Bryden1999,Kley1999,Tatulli2011}. A later study \citep{Carmona2017} of another transitional disk, HD139614, shows also evidence for positive gas surface density gradients inside the submm gap. The $^{12}$CO line profile indicates a very small gap ($<\!1-2$~au), while the optically thinner isotopologues, $^{13}$CO and C$^{18}$O, indicate yet another gas surface density drop at $5-6$~au. At much longer wavelengths and radii larger than 10~au, ALMA provided CO maps and line profiles of gas inside the cavities of several transitional disks \citep[SR21, HD135344B, DoAr44, IRS48][]{vanderMarel2016}. These observations are consistent with remnant gas inside the cavities, but strong supression of the gas surface density by factors of 100 to 10,000 compared to an extension of the profile from the outer disk; the locations of gas density drops are up to a factor two smaller than the submm dust cavities. In some cases, two steps in the surface density profile are required and some of these cases can equally well be fitted by a positive surface density gradient connecting the gap with the outer disk. The spectral resolution of CO ro-vib studies with CRIRES allows to probe gas surface density profiles much closer to the star than ALMA CO rotational studies.

\begin{figure}[!ht]
\begin{center}	
\vspace*{-5mm}
\resizebox{1\hsize}{!}{
	\includegraphics[width=0.46\textwidth]{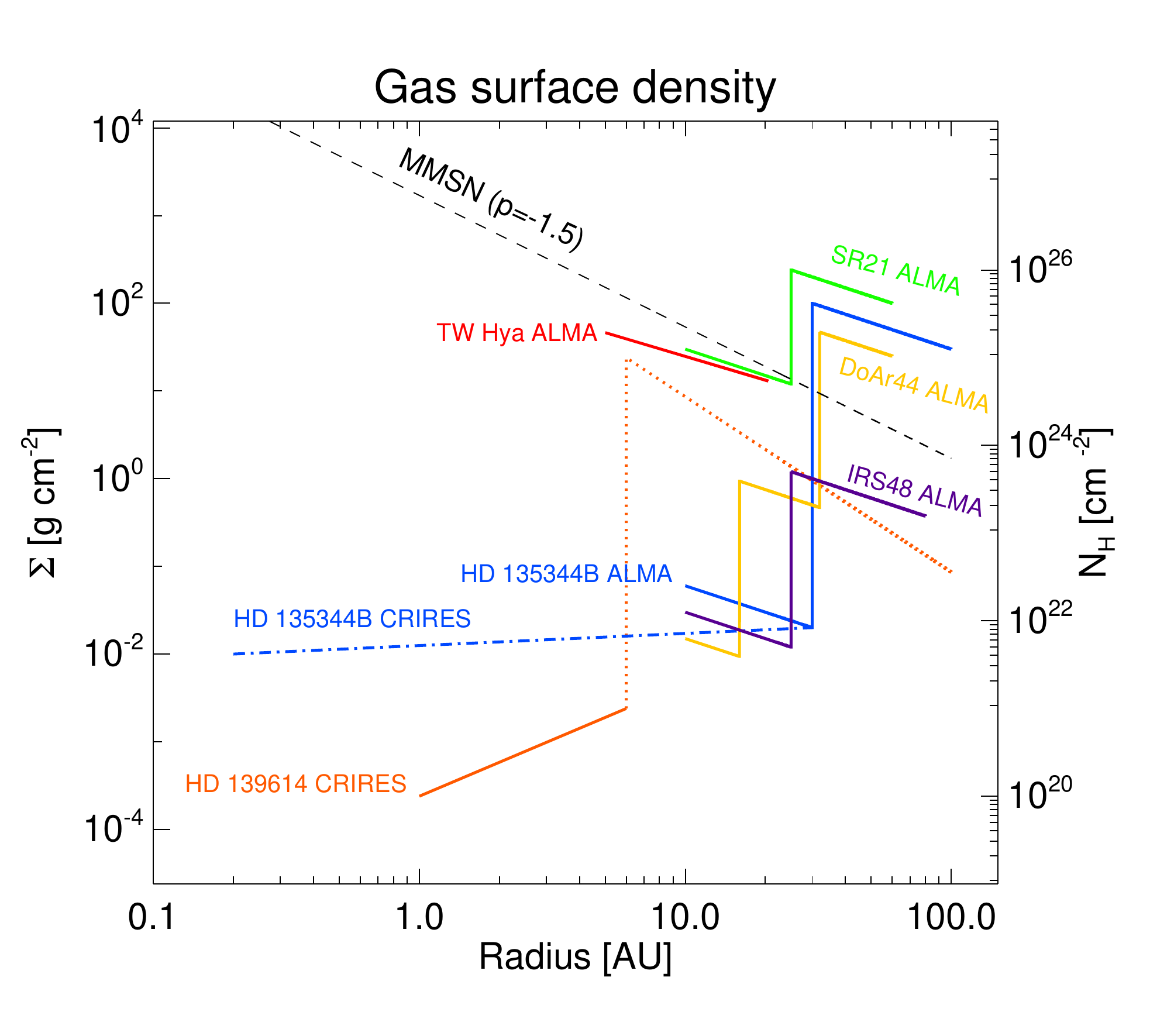}	
	\includegraphics[width=0.54\textwidth]{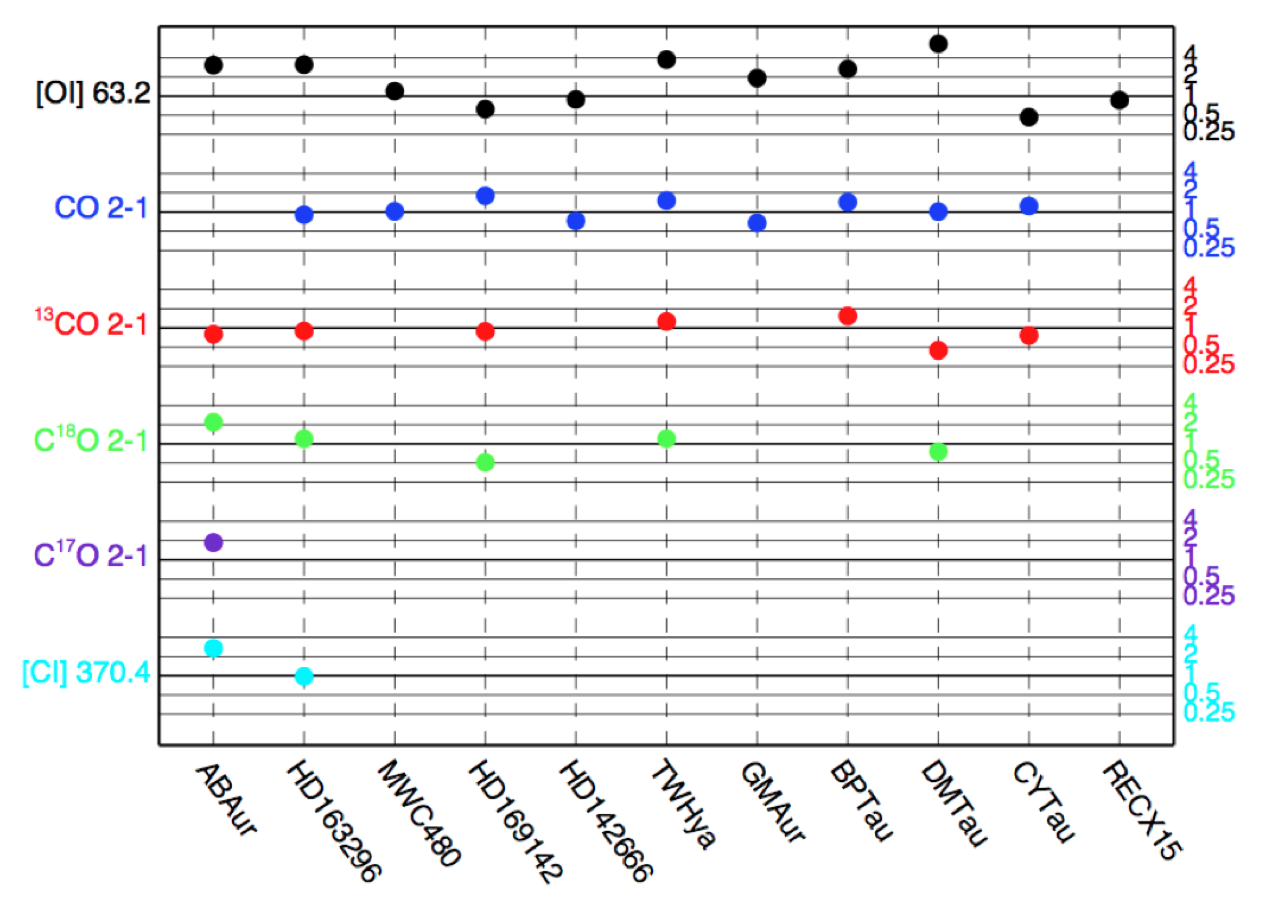}	
	}
	\caption{Left: Gas surface density profiles extracted from the literature (CRIRES and ALMA data --- see references in text); the dashed black line provides the Minimum Mass Solar Nebula profile for comparison; the dotted line for HD139614 is an extension based on SED continuum fitting and assuming a gas-to-dust mass ratio of 100. Right: Ratio between observed and model line fluxes for a selection of far-IR to submm emission lines in a subselection of DIANA disk models.
	}\label{fig:DIANA-linefitting-surfdens}
\end{center}
\vspace*{-5mm}
\end{figure}

The absolute column densities in inner disks can be derived from optically thin isotopologues \citep[e.g.\ CO submm lines][]{Bruderer2013}. \citet{Carmona2017} study gas inside 6 au by using $^{12}$CO, $^{13}$CO and C$^{18}$O ro-vibrational lines to limit the gas column densities inside the cavity to $10^{19}-10^{21}$~cm$^{-2}$. The best fitting gas temperature inside the gas cavity ranges from 400-600~K. Using the very rare isotopologue $^{13}$C$^{18}$O, \citet{Zhang2017} derived the gas surface density power law in the disk of TW Hya between $\sim\!6$ and 20.5~au. The absolute value of surface density is close to that postulated for the Minimum Mass Solar Nebula \citep[MMSN,][]{Hayashi1981}, but the power law exponent is a bit shallower ($-0.9\pm0.4$ compared to $-1.5$). Fig.~\ref{fig:DIANA-linefitting-surfdens} compares the profiles derived so far from CO ro-vib (CRIRES) and CO submm lines (ALMA) with the MMSN. Many of these transitional disks contain less gas column density inside 20~au than the MMSN. The amount of gas depletion compared to the MMSN varies over orders of magnitude. It is however important to keep in mind that the sample shown here is not homogeneous in any respect, but a loose collection of transitional disks with very different central star masses, radiation fields (Spectral Types) and ages.

\section{Chemical evolution in disks: Exoplanet (atmosphere) composition}

\subsection{Element abundances}

The interstellar medium shows very low metal abundances \citep[e.g. from UV gas phase studies of][]{Savage1996} and chemical models of dense interstellar clouds confirm that low electron abundances (metals being the main electron donors) are key to reproducing the observed molecular abundances \citep{Graedel1982}. This is generally understood by an increasing degree of dust condensation \citep{Jenkins2009}. \citet{Voshchinnikov2010} show that this picture of dust being the complement of the observed gas phase interstellar abundances is more complicated and growth and destruction processes in different environments likely occur. Rab et al.\ (in prep) show that the metal abundance in disks is key to reproducing the `average' submm line pattern of T Tauri disks including molecular ions such as HCO$^{+}$ and N$_2$H$^+$. Very low metal abundances such as those used by \citet{Lee1998} (factor $\sim\!100$ below diffuse ISM) are required to achieve a good match of line fluxes as well as the line ratios from those molecular ions.

Besides this metal depletion and its crucial role for molecular abundances, it has been recently suggested that protoplanetary disks have C/O elemental ratios very different from solar. \citet{Bruderer2012} and \citet{Kama2016} find from a combination of CO and [C\,{\sc i}] observations that carbon should be depleted in the surfaces of disks by a factor $\sim\!5-10$. \citet{Favre2013} infer a factor $3-100$ depletion (depending on gas temperature) for CO in the disk around TW Hya from simultaneous fitting of HD and C$^{18}$O. The right panel of Fig.~\ref{fig:DIANA-linefitting-surfdens} shows the result of the fitting of 11 disks within the DIANA project, two of them including existing CO and [C\,{\sc i}] lines. None of these models has a change in carbon and/or oxygen abundance and still they reproduce the observed line fluxes within a factor of a few. It is important to note that several of these models are two-zone models and the radial/vertical extent of the inner disk can have a major impact on the irradiation of the outer disk (shadowing), thus influencing the submm lines of CO and [C\,{\sc i}]. In the DIANA model fits, the inner disk is constrained by simultaneous use of near-IR lines, photometry and if existent interferometry. A more general concern for the interpretation of these submm lines is the impact of changing dust properties (e.g.\ grain size distributions and thus opacities) at the outer disk edge, e.g.\ in the case of radial migration.

\subsection{The role of the iceline locations for gas phase C/O ratios}

The disk midplanes where most of the action of planet formation is happening, is often obscured even at submm wavelengths. In these `dark' regions of the disk, cosmic rays often provide the main ionization source and they can generate secondary UV radiation fields through excitation of H$_2$ \citep{Prasad1983}. This secondary UV field influences photoionisation, -dissociation and -desorption processes in the midplane. \citet{Chaparro2012} show that the field in disk midplanes can be 40 times stronger than previously estimated for dense molecular clouds due to grain growth processes. This affects the ice composition at typical disk lifetimes of $1-3$~Myrs.

\begin{figure}[!ht]
\begin{center}	
\vspace*{-2mm}
	\includegraphics[width=0.325\textwidth]{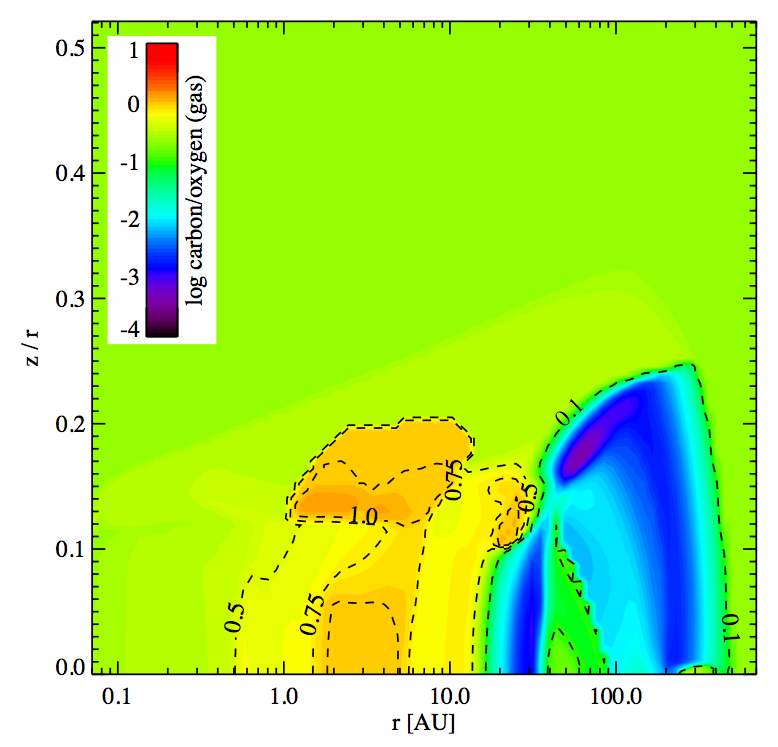}	
        \includegraphics[width=0.325\textwidth]{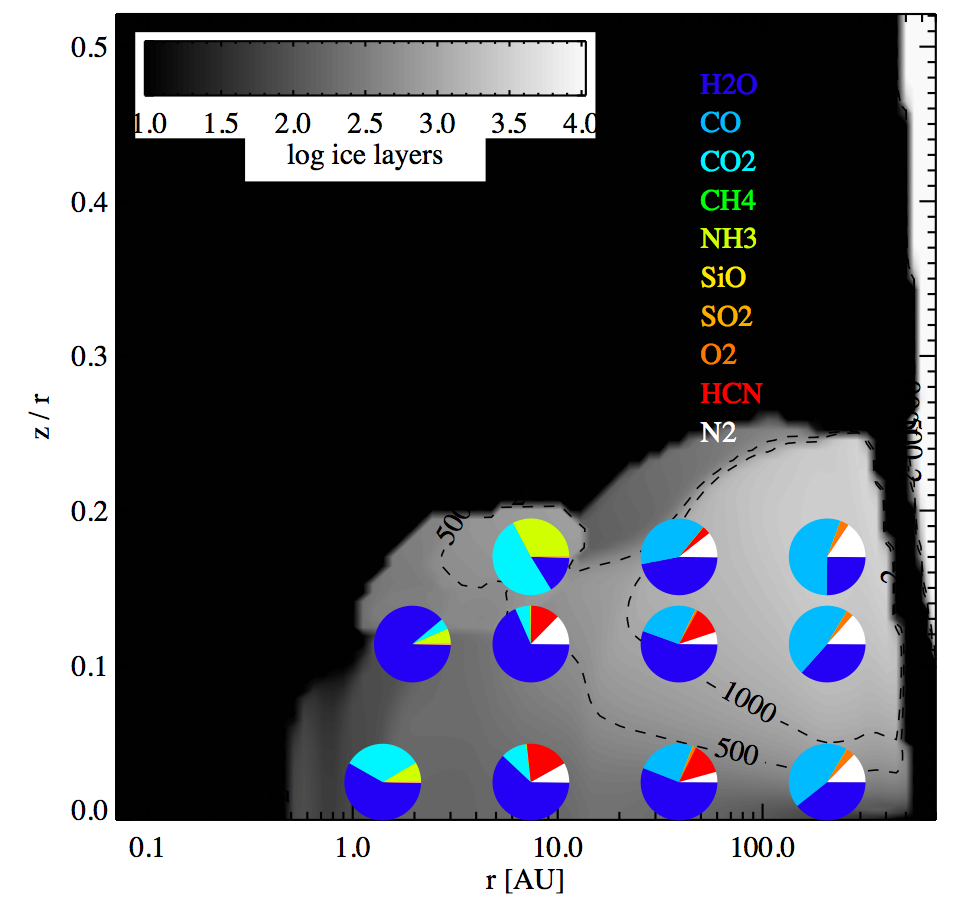}	
	\includegraphics[width=0.34\textwidth]{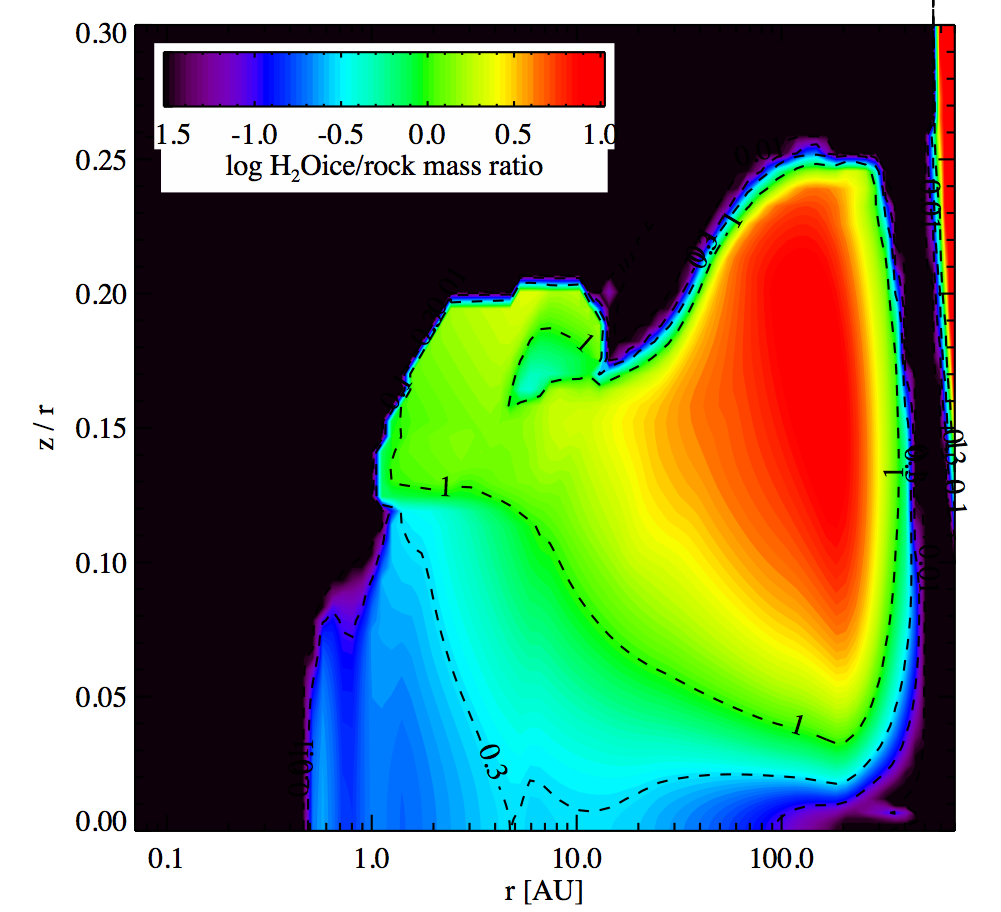}	
	\caption{Results from a typical T Tauri disk model after 1~Myr. Left: C/O ratio in the gas phase. Middle: Number of ice layers with pies that provide ice composition in those locations with the displayed color code. Right: Water ice/rock ratio. Due to settling, the water ice/rock ratio increases above the midplane at larger radii. Note also that the vertical scale is different in the right panel.}.
	\label{fig:TTauri-ices}
\end{center}
\vspace*{-9mm}
\end{figure}

\citet{Oeberg2011} calculate the position of icelines from power law disk models and construct the C/O ratio in various phases using observational abundances of C and O in ices and solids as input. The location of the icelines has a profound impact on the gas phase C/O ratio: Going outwards in the disk and crossing the water, CO$_2$ and CO icelines, the C/O ratio changes from 0.6 to 0.85 and eventually to 1. Later work by \citet{Helling2014} using detailed radiation thermo-chemical disk models and starting with molecular cloud initial conditions shows how the C/O ratio changes as a function of time and distance from the central star. Figure~\ref{fig:TTauri-ices} (left) shows the C/O ratio in the standard T Tauri disk model after 1~Myr. However, it is important to note that the ice composition and thus the gas phase C/O ratio will be affected by the adsorption energies assumed for various molecules. The adsorption energies change with the surface, e.g. NH$_3$ has different adsorption energies on bare grains as compared to a polar ice such as water and this affects the ice composition throughout the disk \citep{Kamp2017}. Also, if metals completely disappear from the gas phase, the chemistry is reset (all elements intitially in atomic form) and grains are small ($0.1~\mu$m), water condensation can be incomplete beyond the snowline \citep{Eistrup2016}.

Figure~\ref{fig:TTauri-ices} (middle panel) shows the ice composition in a standard T Tauri disk model \citep{Woitke2016,Kamp2017} as a function of position in the disk after 1~Myr. This model uses a chemical network with 100 species, molecular cloud initial abundances, very low metal abundances (elements heavier than C, N, O) and the UMIST2012 database (also for ice adsorption energies), and grain size distributions and compositions typical for disks (0.1~$\mu$m$-3$~mm). The dust grains in this model are settled towards the midplane and they can have several hundred to thousands of monolayers of ice. The composition of the ice changes not only as a function of distance, but also with height in the disk. O$_2$ ice --- detected in comet 67P \citep{Rubin2015} --- only forms a notable ice fraction (up to $\sim\!4$\% of total ice, O$_2$/H$_2$O$\!\sim\!6-8$\%) beyond 100~au. Also, the relative fraction of water/CO$_2$ ice is a strong function of position in the disk. Any radial/vertical transport processes can of course alter this composition. \citet{Booth2017} show that growth and radial drift of pebbles can even produce super-solar C/O ratios in giant planets.

\subsection{The ice/solid ratio through the disk midplane}

Ice formation in molecular clouds and also in protoplanetary disks is not an equilibrium condensation process. It requires the presence of dust grain surfaces and will thus be limited by the available surface area (function of grain size distribution) and evolutionary timescale. \citet{Chaparro2012} and \citet{Helling2014} show that the ice formation beyond a few au is not complete at the typical gas lifetime of protoplanetary disks ($1-3$~Myr). 

Figure~\ref{fig:TTauri-ices} (right) shows the water ice/rock ratio for the same T Tauri standard disk model discussed previously. The typical ratio in the comet forming region ($10-20$~au) is $\sim\!0.3$. Assuming that from a comet with such an ice/rock ratio, ices evaporate, this would lead to a gas/dust mass ratio of $\sim\!0.3$. This is  similar to the gas/dust mass ratio measured in comet 67P \citep[dust/gas ratio $4\!\pm2$,][]{Rotundi2015}. In comparison, \citet{Lodders2003} proposed a very high water ice/rock ratio of 1.17 based on equilibrium condensation for a Solar system element composition and 2 for Solar composition gas. 

\citet{Eistrup2016} show that more detailed grain-surface chemistry (beyond adsorption and thermal/non-thermal desorption) will lead to higher CO$_2$/H$_2$O ice ratios compared to the simpler chemistry. They also show that the abundance of O$_2$ inside 20~au and also O$_2$ ice beyond 20 au is largely enhanced if the chemistry in the disk is completely reset (elements initially in atomic form).
  
\section{Outlook}

The future for understanding planet formation in disks looks bright with ALMA and VLT/SPHERE being operational simultaneously to characterize disk substructure and its relation to planet formation. In 2019, we will have JWST studying water and many organic molecules in the terrestrial planet forming region. CRIRES+ can provide after commissioning in 2018 surface density profiles of gas inside cavities for many more objects and shed light on disk dispersal processes and timescales. SOFIA/HIRMES (approved 3rd generation instrument) will study HD and water ice in several bright disks (very massive or very close disks in HD and mostly ices in Herbig disks) and eventually SPICA (proposed JAXA/ESA mission) will do HD and ices in large unbiased samples of T Tauri disks. Over the next years, the results both of observed disk chemical composition as well as kinetic chemistry modeling need to be fed back into planet formation and planet population synthesis models such as those of \citet{Ali-Dib2014} and \citet{Mordasini2016}. Here, a full treatments of dust migration and chemical modelling will be required to build a full picture of the planet formation process. 

\subsection*{Acknowledgments}

The authors acknowledge funding from the EU FP7-2011 under Grant Agreement no. 284405. JDI gratefully acknowledges support from the DISCSIM project, grant agreement 341137, funded by the European Research Council under ERC-2013-ADG. IK acknowledges financial support from Grants-in-Aid for Scientific Research 25108004 for attending the AKARI conference and a work visit to the Tokyo Institute of Technology. Astrophysics at Queen's University Belfast is supported by a grant from the STFC (ST/P000312/1).




\bibliographystyle{aa}
\bibliography{references}


\end{document}